# Distributed MAC Strategy for Exploiting Multi-user Diversity in Multi-rate IEEE 802.11 Wireless LANs

Da Rui Chen and Ying Jun (Angela) Zhang, *Member, IEEE*

**Abstract**—Fast rate adaptation has been established as an effective way to improve the PHY-layer raw date rate of wireless networks. However, within the current IEEE 802.11 legacy, MAC-layer throughput is dominated by users with the lowest data rates, resulting in underutilization of bandwidth. In this paper, we propose and analyze a novel distributed MAC strategy, referred to as Rate-aware DCF (R-DCF), to leverage the potential of rate adaptation in IEEE 802.11 WLANs. The key feature of R-DCF is that by introducing different mini slots according to the instantaneous channel conditions, only contending stations with the highest data rate can access the channel. In this way, the R-DCF protocol not only exploits multi-user diversity in a fully distributed manner but also reduces the loss of throughput due to collisions. Through analysis, we develop an analytical model to derive the throughput of R-DCF in general multi-rate WLANs. Using the analytical model we investigate the performance of R-DCF protocol in various network settings with different rate adaptation strategies and channel variations. Based on the analysis, we further derive the maximal throughput achievable by R-DCF. For practical implementation, an offline adaptive backoff method is developed to achieve a close-to-optimal performance at low runtime complexity. The superiority of R-DCF is proven via extensive analyses and simulations.

**Index Terms**—IEEE 802.11 WLANs, medium access control (MAC), multi-user diversity, performance anomaly, rate adaptation.

——————————— ◆ ———————————

## 1 INTRODUCTION

WIDELY adopted at home, offices, and hot spots, IEEE 802.11 WLANs (wireless local area networks) are expected to provide services parallel to its wired counterpart in near future [1]. To address this requirement, fast rate adaptation, which is capable of drastically enhancing the PHY-layer raw data rate, holds significant promise [2]. Various rate adaptation algorithms have been proposed for WLANs in recent years, targeting at selecting the most suitable transmission rates according to stations' time-varying channel conditions [3], [4], [5], [6], [7]. With rate adaptation, it is common in today's WLANs (e.g. IEEE 802.11a/b/g) that multiple data rates coexist in the network. Unfosion rates, leading to unexpected performance degradation of high-rate stations. This phenomenon, referred to as performance anomaly [8], is due to the fact that current MAC protocol implicitly guarantees throughput fairness to all the users regardless of their transmission rates. Consequently, low rate links occupy much more airtime than high rate links, if the packet sizes are the same. Besides, whenever collision happens, the airtime wasted is typically determined by the lowest transmission rate involved in the collision. It is therefore essential to re-design the MAC strategy for multi-rate WLANs to leverage the advantage of fast rate adaptation and achieve an overall high network throughput.

### 1.1 Related Work
Various schemes have been proposed in recent years to mitigate the performance anomaly and increase the total system throughput of multi-rate WLANs [9], [11], [15], [16], [17]. Bruno *et al* [9] apply the dynamic backoff method [11] to approach the throughput limit of multi-rate WLANs. However, due to the existence of the low-rate transmissions, the throughput limit of a multi-rate WLAN is often much lower than its high PHY-layer available transmission rates. Therefore, directly applying dynamic backoff does not effectively improve the system throughput. In [11], the authors present an Opportunistic Auto Rate (OAR) protocol, where a station transmits multiple packets in proportion to its instantaneous data rate. Similar concept is also introduced in IEEE 802.11e [12] through the use of transmission opportunity (TXOP), where equal airtime is allocated to stations with different data rates and multiple frames can be transmitted within the TXOP interval. Accordingly, airtime fairness instead of throughput fairness is achieved. In all the above schemes, however, only the time-domain channel variation (time diversity) is considered in the protocol design. Low-rate stations have the same chance of grabbing the channel as high rate stations.

In multi-rate WLANs, the channel spectrum is more efficiently utilized when high-rate stations transmit. It is therefore natural to opportunistically grant high-rate stations better chances of accessing the channel by differentiating users at the MAC layer. The significant spectrum-efficiency enhancement therefore achieved, often referred to as multiuser diversity [13], [14], has recently attracted extensive research interest. For example, assuming that stations use different but fixed rates, the authors in [15] propose a remedy scheme to differentiate stations' long-term channel access opportunities by using different backoff parameters according to their data rates. However, due to the rapid fluctuating nature of the wireless channel, it is very common for a mobile station to frequently adapt its transmission rate during one session. Consequently, the long-term backoff-based remedy scheme becomes ineffective when fast rate adaptation is taken into account. In [16] the authors develop an Opportunistic Scheduling and Auto-Rate (OSAR) protocol, where the transmitter probes the channels of multiple intended receivers through RTS/CTS exchange and then choose the best one. Nevertheless, in many applications data frames in a station's buffer are generally targeted to a fixed destination due to service burst, which implies that there is usually only one potential candidate receiver. Hence, OSAR can not effectively gain multi-user diversity in general situations. In [17], the authors propose a User-aware Rate Adaptive Control (UARAC) scheme, where the stations reply CTS packets with a probability that is a function of the channel conditions estimated from RTS packets. The computational complexity, however, is prohibitively high. In addition, in all the above schemes, though low-rate stations no longer occupy excessive air time, the collision cost is still dominated by the lowest data rate, resulting in an unnecessary waste of bandwidth.

## 1.2 Contributions

In this paper we propose and analyze a novel MAC strategy, referred to as R-DCF (Rate-aware Distributed Coordination Function), to exploit multi-user diversity as well as time diversity in a fully distributed manner in IEEE 802.11 DCF-based WLANs. By dynamically adjusting a station's channel access priority through the use of different mini slots, the R-DCF protocol significantly enhances the channel utilization and drastically improves the network throughput. In particular, our contributions include the following:

- The proposal of R-DCF to exploit both multi-user diversity and time diversity in multi-rate WLANs.
- The development of an analytical model to evaluate the throughput performance of the proposed R-DCF scheme in general multi-rate WLANs.
- The analysis and derivation of throughput upper bound of R-DCF for both basic access and RTS/CTS access mode.
- The development of an offline adaptive backoff method to approach the throughput limit of R-DCF with low runtime complexity.

Unlike many multi-rate WLAN analyses and enhancement schemes which assume that stations use fixed data rates for all their packet transmissions [9], [11], [15], R-DCF applies to practical WLANs with various PHY-layer rate-adaptation algorithms, including both fast-rate-adaptation and slow-rate-adaptation schemes. In contrast to existing multi-user diversity-based schemes [15], [16], [17], the R-DCF protocol not only opportunistically favors high rate users but also dramatically dimin-ishes the collision cost by reducing collision probability and avoiding collisions between low-rate and high-rate stations. Different from most enhancement schemes which require significant changes of the well-established DCF protocol [9], [15], [16], [17], the R-DCF protocol requires only a minimal amendment of the standard DCF. Hence, it can be easily implemented in any IEEE 802.11-compliant products. Finally, we conduct in-depth analyses and extensive simulations to verify the superiority of the proposed protocol.

## 1.3 Organization

The rest of the paper is organized as follows. In Section 2 we introduce the standard distributed contention-based MAC protocol in IEEE 802.11 WLANs. The proposed R-DCF protocol is described in Section 3. In Section 4, we theoretically analyze the throughput performance of R-DCF. The performance of R-DCF is evaluated numerically in Section 5. Based on the analytical model, in Section 6 we investigate the maximal throughput that can be achieved by R-DCF and propose an offline adaptive backoff method to approach the maximal throughput. Finally, Section 7 concludes this paper.

## 2   MAC PROTOCOL IN IEEE 802.11 WLANS

### 2.1 IEEE 802.11 DCF

In the IEEE 802.11 protocol, the fundamental medium access mechanism is referred to as distributed coordination function (DCF), which is a random access protocol based on carrier sense multiple access with collision avoidance (CSMA/CA). DCF specifies two access mechanisms, namely the default basic access mechanism and an optional RTS/CTS mechanism. In the basic access scheme, a station with new packet to transmit first monitors the channel. If the channel is idle for a period of time equal to a Distributed Inter-frame Space (DIFS), the station start transmission immediately after this DIFS time. Otherwise, the station continues to motinor the channel until the channel is sensed idle for a DIFS. At this point, the station selects a random backoff time and defers transmission. DCF adopts a binary exponential backoff policy. The backoff time is uniformly chosen from the interval $[0, CW-1]$, where $CW$ denotes the contention window size. At the first transmission attempt, $CW$ is set equal to $CW_{min}$, which is referred to as minimum contention window size. After each unsuccessful transmission, $CW$ is doubled, up to a maximum value $CW_{max}$. The backoff time is decremented as long as the channel is sensed idle, suspended when a transmission is detected on the channel, and reactivated when the channel becomes idle for more than a DIFS time. The station transmits immediately when the backoff time counter reaches zero. When the packet is received successfully, the receiving station will send an ACK frame to the sender after a Short Inter-frame Space (SIFS) and the sender will reset its CW to $CW_{min}$.

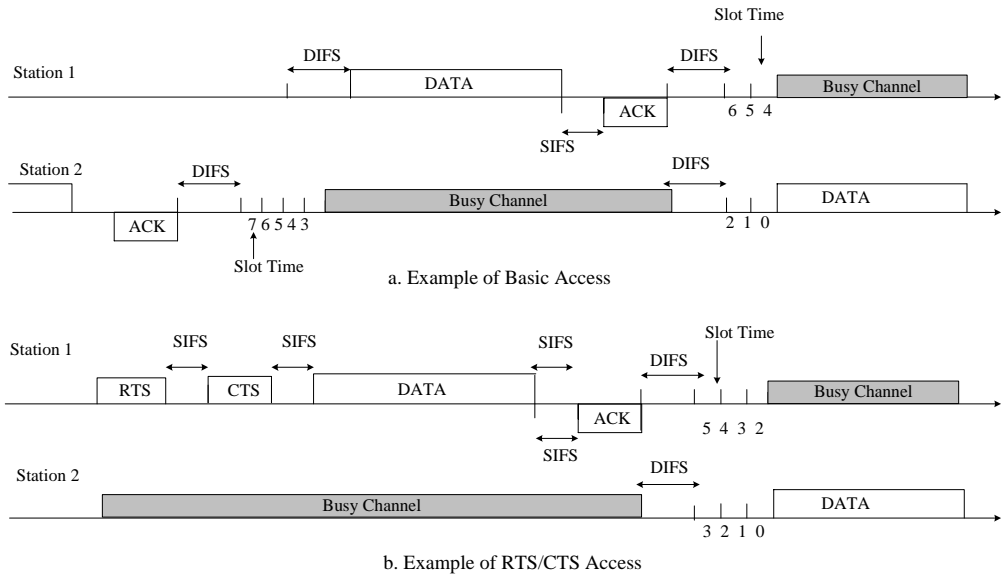

Fig. 1 Basic access and RTS/CTS mechanism in DCF.

In the RTS/CTS mechanism, when a station has a packet to transmit, instead of transmit data packet immediately after winning channel contention, it transmits a short frame called Request-to-Send (RTS). When the receiving station detects the RTS frame, it replies, after a SIFS, with a clear-to-send (CTS) frame. The sender is allowed to transmit only if the CTS frame is correctly received. The RTS/CTS mechanism is effective in terms of system performance especially when large packets are transmitted, as it reduces the length of frames involved in contention period. Fig. 1 illustrates the operation of both the basic access and RTS/CTS mechanism.

## 2.2 Performance Anomaly of DCF

In DCF, each station has the same long-term channel access opportunity because of the same backoff rules imposed on every station. Therefore, throughput fairness is implicitly guaranteed. However, in a multi-rate environment, this is the root cause of low efficiency. Consider a WLAN where two mobile stations transmit equal-size data packets with data rates 6 Mbps and 54 Mbps, respectively. Due to the equal access probability granted to the two stations in DCF, the 6 Mbps-station will occupy as much as 9 times more airtime than the 54 Mbps-station does in order to transmit the same data packet. Consequently, much precious airtime is consumed by the low rate transmissions. Besides, whenever collision happens among the low and high rate stations, the collision duration is always dominated by the lower rate in basic access mode. As a result, the system throughput of this WLAN is only around 6 Mbps.

To better utilize good channel conditions, airtime fairness rather than throughput fairness is recommended as the objective of resource allocation in multi-rate WLANs [19], [20]. A straightforward way to achieve equal airtime allocation is to use packet burst through the use of TXOP as defined in 802.11e. The burst length (TXOP length) is the same for all the users regardless of their data rates. Because every station has the same channel access opportunity in DCF, equal airtime allocation is achieved by each individual station via the same granted TXOP length. Nevertheless, using TXOP helps to improve system spectrum efficiency only when high rate stations win channel contention. As to collision, neither collision probability nor duration can it reduce. Hence solely using TXOP based on original DCF improves system performance only to a limited extent. In the next section, we present the Rate-aware DCF to fully exploit the time-varying nature of the wireless channel in general multi-user, multi-rate WLANs.

## 3 PROPOSED RATE-AWARE DCF FRAMEWORK

In this paper we consider an IEEE 802.11 DCF-based WLAN, where a number of stations operate in either the ad hoc or the infrastructure mode. With rate adaptation, each station dynamically adapts its transmission rate according to the instantaneous channel condition. Assume that there are in total $M$ possible transmission rates denoted by $R_1, R_2, \cdots, R_M$. Without the loss of generality, let $R_1 < R_2 < \cdots < R_M$, and we say a station is in mode $m$ if its current transmission rate is $R_m$.

To exploit multi-user diversity, the proposed R-DCF protocol grants a higher priority to higher-rate links by adding different additional waiting time before each packet transmission. Specifically, when the backoff counter of a mode-$m$ station counts down to zero, the station will wait for an additional duration of time

$$T_{mini,m} = (M - m) \times (\sigma / M) \qquad (1)$$

instead of transmitting immediately as in original DCF. In (1), $\sigma$ denotes the length of an idle slot in our system and $\sigma / M$ is referred to as a *mini slot*. Note that the larger the transmission rate, the shorter the additional waiting time. When several stations' backoff counters count down to zero simultaneously, the highest-rate stations have the shortest waiting time and will initiate transmission after waiting for a period of time specified by (1). The other lower-rate contending stations have longer waiting times and will observe the transmission of the highest rate stations. Consequently, they cannot transmit according to the IEEE 802.11 carrier-sensing strategy. In our scheme, these stations are forced to backoff as if collision had happened. In the rest of the paper, we refer to this kind of "collision" as virtual collision. Meanwhile, all the other stations will detect the transmission on the channel and freeze their backoff counters as in original DCF. Note that this rate-aware prioritization is applicable to both the basic access and RTS/CTS access modes.

In addition to multi-user diversity, R-DCF makes full use of the favorable channel condition by allowing multiple frames to be sent consecutively after a station wins the channel contention. Specifically, the number of consecutively transmitted packets is proportional to the station's instantaneous data rate. Consequently, high rate stations can transmit more packets and favorable channel condition is better utilized.

The salient feature of R-DCF protocol is that it effectively reduces *actual* collisions and opportunistically turns collision into successful transmission at the highest available rate. In original DCF, all stations that count down to zero at the same time will collide with each other. In contrast in R-DCF, only the highest-rate station(s) will attempt transmission. Hence, no collision will occur if there is no more than one station enjoying the same highest rate. In addition, multiple frames can be transmitted at the highest available rate. Furthermore, even if collision occurs, it only involves transmissions at the highest data rate, which implies that the collision duration is drastically reduced.

Fig. 2 shows one illustrative example of this rate-aware prioritization in an IEEE 802.11a WLAN [26]. According to 802.11a protocol, the possible data rates include (6, 9, 12, 18, 24, 36, 48, 54) Mbps. At time $t_1$, STA (station) 1, 2 and 3's backoff counters reach zero, while STA 4 is in the middle of its backoff process. Assume that the data rates of STA 1, 2, 3 are 48Mbps, 36Mbps, and 36Mbps, respectively. Since 48Mbps and 36Mbps are the second and third highest rate among all the 8 modes, the waiting times for the three stations are equal to $1 \times (\sigma/8)$, $2 \times (\sigma/8)$, $2 \times (\sigma/8)$, respectively. Thus, STA 1 can successfully send its packet at time $t_2$ after waiting for $1 \times (\sigma/8)$, while STA 2 and 3 will observe the transmission of STA 1 and backoff. On the other hand, STA 4 freezes its backoff counter starting from $t_2$. Obviously, a successful transmission at the currently highest rate (48 Mbps) is obtained instead of a collision at the lowest rate (36 Mbps) as in original DCF.

In practice, the mini slot $\sigma/M$ should be long enough for the stations to detect the transmission attempts of high-rate links. Thus, in our system, we propose to set the time slot $\sigma$ to be $M$ times as long as an ordinary idle slot in the original IEEE 802.11 standard. With this modification, the mini slot is as long as the idle slot in original 802.11 and is long enough for the stations to sense the channel. As will be shown in our performance evaluations later, the additional overhead introduced by enlarging the slot length hardly affects network throughput, since the effect of slot size on throughput is marginal [20],

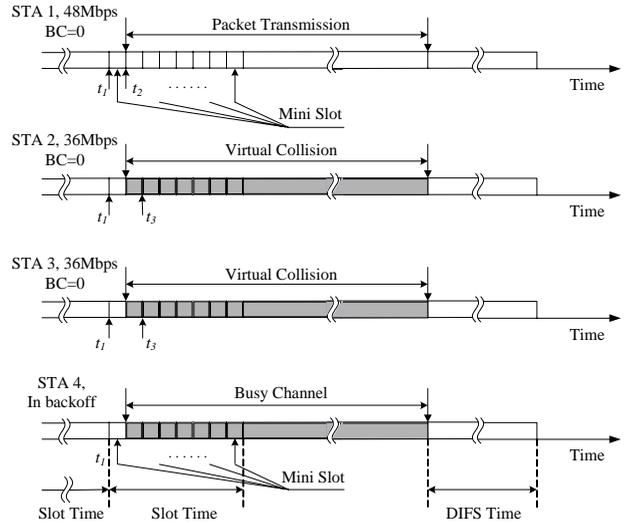

Fig. 2 Prioritization of R-DCF via additional mini slots.

[21].

Note that the only modifications of R-DCF compared with DCF are a larger slot length and the rate-aware deferred transmission using the mini slots. All these functionalities are provided through the carrier-sensing and backoff procedure as basic features in DCF. Therefore, R-DCF can be easily implemented in any IEEE 802.11-compliant product with marginal amendment. In addition, most of existing enhancing schemes for DCF can be applied into R-DCF with almost no changes. For example, R-DCF can be deployed in the RTS/CTS mode to further reduce collision duration and combat the "hidden node" problem [22]. Existing studies on dynamic backoff algorithms [10], [23], [24], [25] can be adopted in R-DCF to achieve a throughput upperbound at runtime. In the following sections, we theoretically analyze the throughput performance of R-DCF in a general multi-rate environment and present enhancing schemes based on the proposed R-DCF framework.

## 4 THEORETICAL ANALYSIS OF R-DCF

In this section, we analyze the performance of the proposed R-DCF protocol. We will first introduce the PHY-layer rate adaptation models. Then, saturation throughput will be analyzed based on a Markov chain model.

### 4.1 Modeling the PHY-layer Rate Adaptation

The general idea of rate adaptation is that a high-level modulation scheme requires a higher signal-to-noise ratio (SNR) to obtain the same specified bit error rate (BER) or packet error rate (PER) requirement in respect to low-level modulation schemes. Hence, it is sufficient to model the PHY-layer rate adaptation by studying the channel variations. Finite-state Markov chain models are widely proposed to represent the time-varying behavior of fading channels [6], [27], [28]. Specifically, the received SNR are partitioned into a finite number of states and each state corresponds to different channel quality, which determines the achievable data rate. Let $\gamma_0, \gamma_1, \cdots, \gamma_{M-1}, \gamma_M$ be received SNR thresholds in increasing order with $\gamma_0 = 0$ and $\gamma_M = +\infty$. The channel is said to be in state $m$ and rate $R_m$ is chosen if $\gamma \in [\gamma_{m-1}, \gamma_m)$. The SNR thresholds are carefully selected according to the probability density function (PDF) of SNR, the prescribed BER/PER, and the coding and modulation schemes adopted. Detailed partitioning method can be found in [27], [28]. Once the number of states and the corresponding SNR thresholds have been determined, the stationary distribution of state $m$ seen by station $n$, which is also the probability of rate $R_m$ being adopted by station $n$, can be calculated as:

$$P_{n,m} = \int_{\gamma_{m-1}}^{\gamma_m} p_n(\gamma) d\gamma , \qquad (2)$$

where $p_n(\gamma)$ is the PDF of the SNR of station $n$.

For an arbitrary WLAN, generally channel statistics are different for different stations depending on their location, mobility, surrounding environment and so on. With fast rate adaptation, a station selects transmission rate based on its instantaneous SNR, which is determined by multi-path fading and varies rapidly from packet to packet. Hence, users are usually heterogeneous in terms of their stationary distribution of transmission rates. Correspondingly, in our analysis $P_{n,m}$'s are considered to be different to model a general multi-rate WLAN. As a special case, when channel fadings are statistically independent and identical to all stations, which means $p_n(\gamma)$ is the same for each individual station, different stations will have the same probability $P_m$ to use rate $R_m$. In this case, stations become homogeneous with the same transmission rate distribution $(P_1, \cdots, P_M)$. On the other hand, in some rate adaptation schemes [8], transmission rates of each station are selected based on the long-term average SNR, which is mainly determined by path loss and shadowing effect, and varies relatively slowly compared with the duration of a session. As an extreme case of such slow-adaptation strategies, we also study in this paper the fixed-rate WLAN scenario, where the data rate used by a station is fixed once chosen. For each individual station, it adopts a certain transmission with probability one and all others zero.

The above two scenarios, namely, the homogeneous-user WLAN and the fixed-rate WLAN represent two extremes in terms of stations' rate distributions. Specifically, in the homogeneous-users case, we investigate the system performance without bias introduced by different users' rate distributions; in the fixed-rate case, we can observe the system's behavior when stations are most affected by different rate distributions. Therefore, in our later numerical study we will investigate the performance of the proposed R-DCF protocol in these two settings to infer the performance of a multi-rate WLAN with more general rate distribution scenarios. In addition, in the following analysis we assume the frame transmissions are almost error free, since we can always gurantee a sufficiently small SNR through rate adaptation.

## 4.2 Throughput Analysis

Throughout our analysis, we assume there are $N$ stations in the network and each station always has a packet available for transmission (saturation condition). Similar to the assumptions in [10], [11], [15], [20], [21], we assume a fully connected network, in which collision occurs when two or more stations transmit simultaneously. Note that we make these assumptions for the tractability of the theoretical analysis. Our scheme is applicable to general wireless networks without such constraints.

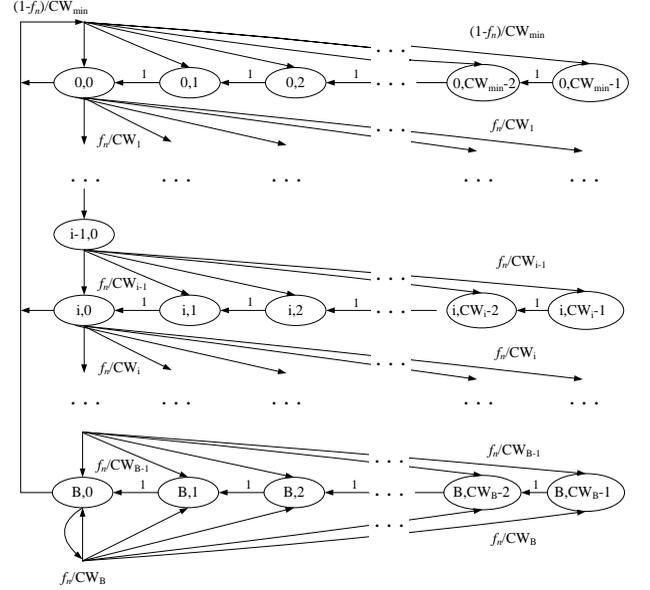

Fig. 3 Markov chain model for backoff process.

In our analysis, we adopt a bi-dimensional Markov chain model based on Bianchi's well-known work [20]. However, the derivations of transmission, collision probability and throughput expression are very different from that in [20], due to the priorities associated with different transmission rates in R-DCF.

### 4.2.1 Zero Backoff Counter Probability and Contention Failure Probability

For a given station, let $b(t)$ be the stochastic process representing the backoff time counter value for a given station at discrete time $t$. Define generic slot to be the variable time interval between two consecutive backoff time counter decrements. A station will increase its contention window after it encounters a collision. The window size is calculated as $CW_i = r^i CW_{min}$, where $i \in [0, B]$ is referred to as the "backoff stage", $B$ the maximum backoff stage, $CW_{min}$ the minimum contention window and $r$ the backoff exponent. The backoff counter value is uniformly chosen in $[0, CW_i - 1]$. Let $s(t)$ be the stochastic process representing the backoff stage $(0, \ldots, B)$ of the station at time $t$. Then, the stochastic process $\{s(t), b(t)\}$ is a discrete Markov chain, shown in Fig. 3.

In R-DCF, a station may not be able to transmit when its backoff counter reaches zero, if there are other backoff counters of higher-rate stations reaching zero at the same time. In this case, we say a virtual collision occurs to the station. Let us define as *candidate station* the stations with zero backoff counter values. For a given station $n$, define $\tau_n$ to be the *zero backoff counter probability*, which is the probability that a station's backoff counter is zero at any given generic slot. Likewise, define $f_n$ to be the *contention failure probability*, which is the probability that a station encounters either actual collision or virtual collision, given its backoff counter is zero. Since generally stations may have different rate distribution, stations that are more likely to initiate transmission at a higher data rate will have better chance to access the channel. Therefore, both $\tau_n$ and $f_n$ are different for different stations. In steady state, for a tagged station, $\tau_n$ and $f_n$ are independent of the number of retransmission and instantaneous transmission rate of this station.

Given the above definitions and assumptions, the non-zero one-step transition probabilities in the Markov chain are

$$\begin{cases} P\{i,k \mid i,k+1\} = 1 & 0 \le k \le CW_i - 2, \quad 0 \le i \le B \\ P\{0,k \mid i,0\} = (1-f_n)/CW_{min} & 0 \le k \le CW_{min}-1, \; 0 \le i \le B \\ P\{i,k \mid i-1,0\} = f_n/CW_i & 0 \le k \le CW_i-1, \quad 1 \le i \le B \\ P\{B,k \mid B,0\} = f_n/CW_B & 0 \le k \le CW_B - 1 \end{cases} \quad (3)$$

Let $b_{i,k} = \lim_{t \to \infty} P\{s(t)=i, b(t)=k\}$ be the stationary distribution of the chain, we have

$$\begin{cases} b_{i,0} = f_n^i b_{0,0} & 0 < i < B \\ b_{B,0} = \dfrac{f_n^B}{1-f_n} b_{0,0} \end{cases} \quad (4)$$

Using the chain regularity, for each $1 \le k \le CW_i - 1$ it is

$$b_{i,k} = \frac{CW_i - k}{CW_i} \cdot \begin{cases} (1-f_n)\sum_{j=0}^{B} b_{j,0} & i = 0 \\ f_n \cdot b_{i-1,0} & 0 < i < B \\ f_n \cdot (b_{B-1,0} + b_{B,0}) & i = B \end{cases} \quad (5)$$

Using the normalization condition $\sum_{i=0}^{B} \sum_{k=0}^{CW_i-1} b_{i,k} = 1$, we can relate all the unknowns to $b_{0,0}$ and $f_n$:

$$b_{0,0} = \frac{2(1-rf_n)(1-f_n)}{(1-rf_n)(CW_{min}+1)+CW_{min}f_n(r-1)\left[1-(rf_n)^B\right]} . \quad (6)$$

Finally, the zero backoff counter probability $\tau_n$ is calculated as

$$\begin{aligned} \tau_n &= \sum_{i=0}^{B} b_{i,0} \\ &= \frac{2(1-rf_n)}{(1-rf_n)(CW_{min}+1)+CW_{min}f_n(r-1)\left[1-(rf_n)^B\right]} . \end{aligned} \quad (7)$$

In (7), $\tau_n$ depends on the contention failure probability $f_n$, which is still unknown. To facilitate our later analysis, we first introduce the following notations. Define $T^N = \{\tau_1, ..., \tau_N\}$ as the set of $\tau_n$ for all the station. Define $T_k^N$ as a $k$-subset of $T^N$, which contains $k$ elements in $T^N$, and $\overline{T_k^N} = \{\tau_i \mid \tau_i \in T^N \; \& \; \tau_i \notin T_k^N\}$, the complement of $T_k^N$ with respect to $T^N$. Denote each $k$-subset as $T_{k,j}^N, j \in [1, C_N^k]$, where $C_N^k = N!/(k!(N-k)!)$ is the total number of different $k$-subsets on an $N$-element set. To calculate $f_n$, consider that at any given slot, when station $n$'s backoff counter is zero, the station would have different conditional collision probability, depending on its own transmission rate and transmission rates adopted by other candidate stations. Let $f_{n,m}$ be the contention failure probability seen by a packet transmitted at rate $R_m$. We have

$$f_{n,m} = \sum_{k=1}^{N-1} \left\{ \sum_{j=1}^{C_{N-1}^k} \sum_{\substack{s: \\ \tau_s \in T_{k,j}^{N-1}}} \prod \tau_s \cdot \left[ 1 - \prod_{\substack{s: \\ \tau_s \in T_{k,j}^{N-1}}} \left( \sum_{l=1}^{m-1} P_{s,l} \right) \right] \cdot \prod_{\substack{t: \\ \tau_t \in \overline{T_{k,j}^{N-1}}}} (1-\tau_t) \right\} \quad (8)$$

In (8), $j$ is the index used to enumerate all the $k$-subsets of $T^{N-1}$. For a given $k$-subset $T_{k,j}^{N-1}$, $\tau_s \in T_{k,j}^{N-1}$ represents the transmission probability of stations in this set and $\tau_t \in \overline{T_{k,j}^{N-1}}$ represents that of the other stations. Therefore, $\prod_{s:\tau_s \in T_{k,j}^{N-1}} \tau_s$ means $k$ stations are now candidates and ready to transmit, whereas $\prod_{t:\tau_t \in \overline{T_{k,j}^{N-1}}}(1-\tau_t)$ means the rest stations do not attempt transmission at the very slot. $P_{s,l}$ is the probability station s in the $j$-th $k$-subset chooses to use rate $R_l$. $\sum_{l=1}^{m-1} P_{s,l}$ is the probability that station $s$ chooses transmission rate lower than current rate $R_m$. The interpretation of (8) is that, except for the tagged station $n$, given altogether $k \in [1, N-1]$ candidate stations, the tagged station $n$ transmitting packets at rate $R_m$ will collide actually or virtually, when there exists at least one candidate station whose current transmission rate is higher than or equal to $R_m$. Based on (8), the average contention failure probability $f_n$ for station $n$ can be calculated as

$$f_n = \sum_{m=1}^{M} P_{n,m} \cdot f_{n,m} , \quad (9)$$

where $P_{n,m}$ is the probability that station $n$ uses rate $R_m$. Equation (7), (8) and (9) represent a nonlinear system in two groups of unknowns $\tau_n$ and $f_n$, which can be solved numerically.

### 4.2.2 Saturation Throughput

We define throughput $S$ as the ratio of the expected data payload transmitted during a generic time slot to the expected length of a generic time slot, which can be expressed as

$$S = \frac{\sum_{m=1}^{M} P_{succ,m} L_m}{\sum_{m=1}^{M} P_{succ,m} T_{succ,m} + P_{idle}\sigma + \sum_{m=1}^{M} P_{coll,m} T_{coll,m}} , \quad (10)$$

where $L_m$ is the data payload transmitted at rate $R_m$; $T_{succ,m}$ and $T_{coll,m}$ are the successful transmission time and collision time when stations of rate $R_m$ are involved, respectively; $\sigma$ is the length of slot time; $P_{idle}$ is the channel idle probability; and $P_{succ,m}$ and $P_{coll,m}$ are the successful transmission probability and collision probability at $R_m$, respectively.

To compute $S$, we need to calculate the possible durations of the generic slot and their corresponding probabilities in (10). First, for basic access and RTS/CTS mechanism, $T_s$ and $T_c$ can be easily obtained respectively:

$$\begin{cases} T_{succ,m}^{basic} = (m-1)\sigma/M + T_H + L_m/R_m + SIFS \\ \qquad\qquad + ACK + DIFS \\ T_{coll,m}^{basic} = (m-1)\sigma/M + T_H + L_m/R_m + DIFS \end{cases} \quad (11)$$

$$\begin{cases} T_{succ,m}^{RTS} = (m-1)\sigma/M + RTS + SIFS + CTS + SIFS \\ \qquad\qquad + T_H + L_m/R_m + SIFS + ACK + DIFS \\ T_{coll,m}^{RTS} = (m-1)\sigma/M + RTS + DIFS \end{cases}, \quad (12)$$

where $(m-1)/M \cdot \sigma$ is the additional mini slots introduced by R-DCF, $T_H$ is the packet header transmission time and the acronyms (i.e. ACK, DIFS, SIFS,RTS, CTS) represent the corresponding time duration specified in the standard.

Now we calculate in detail the unknown probabilities in (10). First, the channel idle probability $P_{idle}$ is obtained considering that the channel will be idle only when no station's backoff counter is zero at a given time slot. Thus, we have

$$P_{idle} = \prod_{i=1}^{N} (1 - \tau_n). \quad (13)$$

To have a successful transmission at rate $R_m$, it is sufficient to notice that, given several stations' backoff counter counting down to zero and ready to transmit, only one of them adopts $R_m$ and the others shall use rates lower than $R_m$. Let $P_{succ,m}$ be the probability that a transmission occurring on the channel is successful at transmission rate $R_m$. We have

$$P_{succ,m} = \sum_{k=1}^{N} \left\{ \sum_{j=1}^{C_N^k} \left[ \prod_{\substack{s: \\ \tau_s \in T_{k,j}^N}} \tau_s \left( \sum_{x:} P_{x,m} \cdot \prod_{\substack{u: \\ u \neq s, \tau_u \in T_{k,j}^N}} \left( \sum_{l=1}^{m-1} P_{u,l} \right) \right) \prod_{\substack{t: \\ \tau_t \in \overline{T_{k,j}^N}}} (1 - \tau_t) \right] \right\}. \quad (14)$$

In (14), $j$ is the index used to enumerate in total all the $C_N^k$ $k$-subsets of $T^N$, which represents all the possible combinations of $k$ candidate stations out of total $N$ stations. Similar to (4), for a given $k$-subset $T_{k,j}^N$, $\tau_s \in T_{k,j}^N$ represents the transmission probability of stations in this set and $\tau_t \in \overline{T_{k,j}^N}$ represents that of the other stations. $\prod_{s:\tau_s \in T_{k,j}^N} \tau_s$ means $k$ stations are now candidates and ready to transmit, whereas $\prod_{t:\tau_t \in \overline{T_{k,j}^N}} (1 - \tau_t)$ means the rest stations do not attempt transmission at the very slot. Given $k$ candidate stations, a successful transmission happens at $R_m$ only when one adopts $R_m$ and the others use rate lower than $R_m$, which is $\sum_s P_{s,m} \prod_{u:u \neq s, \tau_u \in T_{k,j}^N} \left( \sum_{l=1}^{m-1} P_{u,l} \right)$.

In R-DCF collision happens only if the following two conditions are satisfied: 1) more than one has zero backoff counter value at a given time slot; 2) among all the candidates, at least two of them adopt the same highest data rate. Let $P_{coll,m}$ be the probability that a transmission is collided at rate $R_m$. We have

$$P_{coll,m} = \sum_{k=2}^{N} \left\{ \begin{array}{l} \sum_{j=1}^{C_N^k} \prod_{\substack{s: \\ \tau_s \in T_{k,j}^N}} \tau_s \cdot \left[ \sum_{u=2}^{k} \sum_{v=1}^{C_k^u} \prod_{\substack{x: \\ x \in S_{u,v}^k}} P_{x,m} \cdot \prod_{\substack{y: \\ y \in \overline{S_{u,v}^k}}} \left( \sum_{l=1}^{m-1} P_{y,l} \right) \right] \cdot \\ \prod_{\substack{t: \\ \tau_t \in \overline{T_{k,j}^N}}} (1 - \tau_t) \end{array} \right\} \quad (15)$$

where $S^k \in \{s \mid \tau_s \in T_{k,j}^N\}$ is the set of index $s$ with size $k$; $S_u^k$ is a subset with $u$ elements of $S^k$, with $S_{u,v}^k$ denoting all

$C_k^u$ different subsets; $\overline{S_{u,v}^k}$ is the complement of $S_{u,v}^k$ with respect to $S^k$. Similar to (14), $j$ is the index used to enumerate all the $k$-subsets of $T^N$. $\tau_s \in T_{k,j}^N$ and $\tau_t \in \overline{T_{k,j}^N}$ represent respectively, the transmission probabilities of stations in a particular $k$-subset $T_{k,j}^N$ and its complement $\overline{T_{k,j}^N}$. $\prod_{s:\tau_s \in T_{k,j}^N} \tau_s \prod_{t:\tau_t \in \overline{T_{k,j}^N}} (1 - \tau_t)$ is the probability of having $k$ candidate stations. Since collision occurs at rate $R_m$, there are at least two candidates adopting $R_m$ and the rest use rate lower than $R_m$. Index $u$ is used to count the number of candidates using rate $R_m$. Given $S_k$ the set of the index of candidates, index $v$ is used to enumerate all the $u$-element subsets of $S_k$. $\prod_{x:x \in S_{u,v}^k} P_{x,m} \prod_{y:y \in \overline{S_{u,v}^k}} \left( \sum_{l=1}^{m-1} P_{y,l} \right)$ means among all $k$ candidates, $u$ of them use $R_m$ and the rest use other lower rates. $\sum_{u=2}^{k} \sum_{v=1}^{C_k^u} \prod_{x:x \in S_{u,v}^k} P_{x,m} \prod_{y:y \in \overline{S_{u,v}^k}} \left( \sum_{l=1}^{m-1} P_{y,l} \right)$ represents the probability that collision happens at $R_m$ given $k$ stations are candidates. Summing up over $k$ and all the subsets of $T^N$, we can get $P_{coll,m}$ as in (15).

Now given the channel model described in 4.1, we can first calculate the stationary distribution of transmission rate for each individual station. Then using the Markov chain model, transmission probability $\tau_n$ can be calculated for each station. Finally, we are able to compute the network throughput as a function of $\tau_n$, given the number of stations $N$, available transmission rate $R_m$, corresponding probability $P_{n,m}$.

## 4.3 Simplified Models

Using equations (7) – (15), we are able calculate throughput of R-DCF for general multi-rate WLANs. As stated earlier, in this paper our numerical studies focus on the homogeneous users in the fast-rate-adaptation WLAN and heterogeneous users in the fixed-rate slow-adaptation WLAN. For easy discussion, we simplify our analysis in this sub-section by using the properties of the two scenarios.

### 4.3.1. Homogeneous-user WLANs

In the homogeneous users' scenario, all the users have the same probability $P_m$ to adopt rate $R_m$, since channel fadings are independent and identical to all users, as specified in Section 4.1. In this case, different users have the same the zero backoff probability $\tau$ and contention failure probability $f$, respectively. Hence, the zero backoff counter probability in (7) becomes the same for all users:

$$\tau = \frac{2(1 - rf)}{(1 - rf)(CW_{\min} + 1) + CW_{\min} f(r-1) \left[ 1 - (rf)^B \right]}. \quad (16)$$

In addition, given $k$ candidate stations, all the $k$-subsets $T_{k,j}^N, j \in [1, C_N^k]$ contain the same $k$ elements of transmission probability $\tau$. Hence $\prod_{s:\tau_s \in T_{k,j}^N} \tau_s$ becomes $\tau^k$; $\prod_{t:\tau_t \in \overline{T_{k,j}^N}} (1 - \tau_t)$ becomes $(1 - \tau)^{N-1-k}$ and the summation over all possible combinations becomes the

TABLE 1
IEEE 802.11A SYSTEM PARAMETERS

| Parameters | Value | Comments |
|---|---|---|
| tSlotTime | 9 μs | Slot Time |
| tSIFSTime | 16 μs | SIFS Time |
| tDIFSTime | 34 μs | DIFS Time |
| MAC Header | 224 bits | MAC Header Size |
| PHY Header | 20 μs + 22/6 μs | PHY Header Transmission Time[a] |
| RTS | 160/6 μs + PHY Header | RTS Transmission Time |
| CTS | 112/6 μs + PHY Header | CTS Transmission Time |
| ACK | 112/6 μs + PHY Header | ACK Transmission Time |
| CWmin | 16 | Minimum contention window |
| CWmax | 1024 | Maximum contention window |

[a] *PHY Header Transmission Time includes a 16 μs PLCP preamble, a 4 μs PLCP Signal field, and a 22-bit Service and Trail field transmitted at one of the rates in the BSS basic rate set. We assume 6 Mbps is adopted for header and control frames (i.e. RTS, CTS and ACK frames).*

sum of $C_{N-1}^k$ terms. Therefore, (8), (9) can be simplified as

$$f = \sum_{m=1}^{M} P_m \cdot f_m = \sum_{m=1}^{M} \sum_{k=1}^{N-1} P_m C_{N-1}^k \tau^k (1-\tau)^{N-k-1} \left[ 1 - \left( \sum_{l=1}^{m-1} P_l \right)^k \right]. \quad (17)$$

Similarly, the successful transmission probability $P_{succ,m}$ and collision probability $P_{coll,m}$ at rate $R_m$ calculated in (14) and (15) can be simplified as

$$P_{succ,m} = \sum_{k=1}^{N} C_N^k \tau^k (1-\tau)^{N-k} \left[ k P_m \left( \sum_{i=1}^{m-1} P_i \right)^{k-1} \right] \quad (18)$$

$$P_{coll,m} = \sum_{k=2}^{N} C_N^k \tau^k (1-\tau)^{N-k} \left[ \sum_{i=2}^{k} C_k^i P_m^i \left( \sum_{l=1}^{m-1} P_l \right)^{k-i} \right]. \quad (19)$$

### 4.3.2 Heterogeneous Users in Fixed-rate WLANs

In the fixed-rate WLAN, different stations use different but constant transmission rates. Clearly, stations using the same data rate have the same zero backoff probability $\tau_n$ and contention failure probability $f_n$ respectively, due to the rate-aware prioritization introduced by R-DCF. Categorize stations according to their data rates and let group $m$ represent stations transmitting at rate $R_m$, $N_m$ the number of stations using rate $R_m$. For group $m$, the zero backoff counter probability in (7) becomes

$$\tau_m = \frac{2(1-rf_m)}{(1-rf_m)(CW_{min}+1) + CW_{min}f_m(r-1)\left[1-(rf_m)^B\right]}. \quad (20)$$

For group $m$, a station's transmission attempt will encounter collision either actually or virtually, if stations in similar or higher rate groups have zero backoff counter values in the same slot. Hence, we calculate the contention failure probability for group $m$ as

TABLE 2
MODEL VALIDATIONS

| Access Mode | Basic Access | | | RTS/CTS Access | | |
|---|---|---|---|---|---|---|
| N | S | A | E | S | A | E |
| 5 | 22.79 | 22.80 | 0.44% | 22.15 | 22.22 | 0.32% |
| 10 | 23.80 | 23.97 | 0.71% | 23.20 | 23.47 | 1.2% |
| 15 | 24.46 | 24.63 | 0.72% | 23.87 | 24.17 | 1.2% |
| 20 | 24.91 | 25.09 | 0.72% | 24.36 | 24.67 | 1.3% |
| 25 | 25.30 | 25.44 | 0.55% | 24.77 | 25.05 | 1.1% |
| 30 | 25.61 | 25.73 | 0.18% | 25.09 | 25.37 | 1.1% |
| 35 | 25.85 | 25.97 | 0.46% | 25.38 | 25.63 | 0.98% |
| 40 | 26.10 | 26.19 | 0.34% | 25.63 | 25.87 | 0.93% |
| 45 | 26.34 | 26.39 | 0.19% | 25.04 | 25.09 | 0.2% |
| 50 | 26.51 | 26.56 | 0.19% | 26.06 | 26.27 | 0.80% |

S: *Simulation results*; A: *Analytical results*; E: *Relative error*

$$f_m = 1 - (1-\tau_m)^{N_m-1} \prod_{i=m+1}^{M} (1-\tau_i)^{N_i}. \quad (21)$$

The successful transmission happens at rate $R_m$ if one candidate uses $R_m$ and other candidates use rate lower than $R_m$. Equivalently, this requires no high rate stations attempting transmission, regardless of the low rate stations. Therefore, probability $P_{succ,m}$ is calculated as

$$P_{succ,m} = N_m \tau_m (1-\tau_m)^{N_m-1} \prod_{i=m+1}^{M} (1-\tau_m)^{N_i}. \quad (22)$$

Similarly, collision probability $P_{coll,m}$ at rate $R_m$ can be calculated as

$$P_{coll,m} = \left( \sum_{k=2}^{N_m} C_{N_m}^k \tau_m^k (1-\tau_m)^{N_m-k} \right) \prod_{i=m+1}^{M} (1-\tau_i)^{N_i}. \quad (23)$$

In the next section, we will evaluate the performance of R-DCF in both the fast and slow adaptation cases using the simplified models.

## 5 PERFORMANCE EVALUATION

In this section, we evaluate the throughput performance of R-DCF in various network settings to demonstrate its superiority. In the following experiments, we simulate an IEEE 802.11a system, where there are 8 possible rates, namely 6, 9, 12, 18, 24, 36, 48, and 54 Mbps. Unless otherwise specified, in our numerical study we assume that all the packets have the same size of $L_0 = 2312$ bytes. With packet burst, the number of consecutively transmitted packets $n_m$ is proportional to the data rate, which means $n_m = R_m / R_0$ and $L_m = L_0 R_m / R_0$. Other parameters used to obtain both analytical and simulation results are summarized in Table I.

We first validate the analytical results through numerical simulation. For simplicity, we assume all the rates are equally likely to be used for every user, which means $P_{n,m} = 1/8$ for all $n$ and $m$. In Table II we vary the number of stations $N$ in different experiments. The analytical results are obtained by two steps: 1) given $N$ and backoff parameters, we calculate the zero backoff counter probability $\tau_n$ from (7) − (9). Note that with the same rate distribution $P_{n,m}$ for any $n$, $\tau_n$ becomes the same (denoted by $\tau$) for all the users. 2) Knowing $\tau$, all the unknowns in (10) can be calculated from (11) – (15) and the system throughput is readily obtained by substituting for the unknowns. As shown in the table, analytical and simulation results match well in both the basic access and RTS/CTS modes. The relative errors ($\left| S - A \right| / A$) are always below 1.5%.

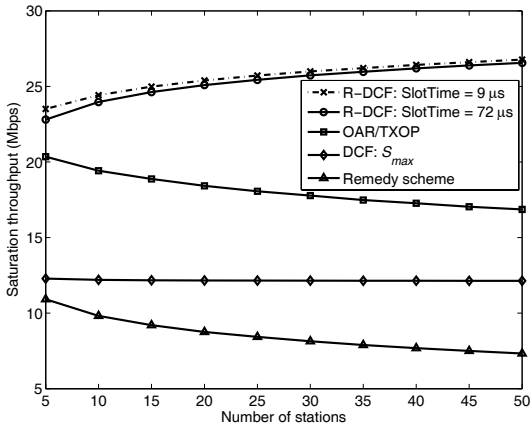

Fig. 4. R-DCF in fast adaptation WLANs: basic-access mode.

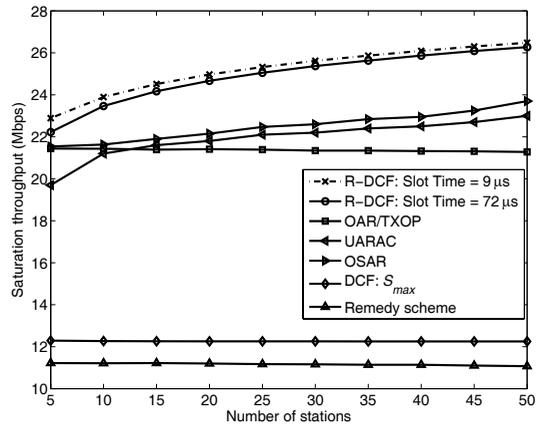

Fig. 5. R-DCF in fast adaptation WLANs: RTS/CTS access mode.

## 5.1 R-DCF with Homogeneous Users

In the homogeneous-user WLAN, each station has the same rate distribution ($P_1, \cdots, P_M$). Unless otherwise specified, we assume for simplicity that $P_{n,m} = 1/8$ for all $n$ and $m$. In Fig. 4, we show the throughput performance of R-DCF in the basic-access mode. For comparison, we also plot 1) the throughput upper-bound, $S_{max}$, of original IEEE 802.11a DCF achieved by dynamic backoff, 2) throughput achieved by using time diversity based schemes (OAR/TXOP), and 3) throughput achieved by differentiating MAC parameters as used in the remedy scheme in [15]. In the OAR/TXOP scheme, we also set the number of consecutively transmitted packets $n_m$ equal to $R_m/R_0$ for fair comparison. In the remedy scheme, we set respectively $CW_{min}$ = 8, 16, 32 and 64 when rates (54 or 48), (36 or 24), (18 or 12), and (9 or 6) Mbps are chosen. From the figure, we can observe that: 1) due to the fast rate adaptation, differentiating MAC parameters does not lead to visible throughput improvement: the remedy scheme performs the worst compared with other schemes. 2) The maximal throughput $S_{max}$ of the original DCF is limited by the existence of low rate transmissions. Hence, directly deploying dynamic backoff does not result in significant throughput improvement. 3) By using time diversity based schemes (OAR/TXOP), system performance is much improved, since more packets are transmitted at high data rates. 4) With the increase of number of users, the proposed R-DCF scheme significantly outperforms the others thanks to its capability of exploiting both multi-user diversity and time diversity gain. With more users present, it is more likely to have high-rate stations involved in successful transmissions and collisions. Thus, throughput is increased and collision cost is reduced. Admittedly, if the number of users grows to infinitely large, the throughput of R-DCF will also decrease due to excessive collisions. However, as shown in the figure, the throughput performance does not degrade in R-DCF even in a network with 50 users. In addition, we also demonstrate that increasing the slot length by $M$ times indeed has marginal impact on the system performance. In the simulations of R-DCF, we assume that the idle slot length equals to 72μs, which is 8 times that of the original 802.11, i.e., 9μs. However, the theoretical throughput with a slot equal to 9 μs is only slightly higher than that of the practical 72μs-slot system.

In Fig. 5, we investigate the performance of R-DCF under RTS/CTS-access mode. In addition to the above schemes compared in basic mode, we also plot the throughput achieved by UARAC and OSAR protocols, which are designed only under RTS/CTS mode. For OSAR we set the number of candidate stations to be four, as suggested in [16]. Similar to the basic-access mode, the maximal throughput of original DCF is limited due to the low rate transmissions. Purely adjusting backoff parameters is not effective in the rate-adaptive WLANs. On the other hand, time diversity based OAR/TXOP scheme sharply increase the system throughput. R-DCF, UARAC and OSAR further outperform the OAR/TXOP scheme, due to the benefit of multi-user diversity. In addition, R-DCF achieves better throughput improvement compared with the UARAC and OSAR schemes. This is because in R-DCF the channel contention is only among the stations adopting the highest data rate, which in fact reduces the actual number of contending stations and consequently the potential collisions. A close observation of Fig. 4 and Fig. 5 reveals that in contrast to original DCF, R-DCF achieves similar performance under both RTS/CTS and basic access modes. This is because through rate-aware prioritization, R-DCF has already avoided most collisions. Meanwhile, the cost of unavoided collisions is small, since collision only happens between high-rate stations. Thus, the advantage of RTS/CTS is in R-DCF not as obvious as in original DCF, when the network is fully connected.

To demonstrate R-DCF's effectiveness in reducing collision, in Fig. 6 and Fig. 7 we plot, respectively, the collision probability and the collision cost for R-DCF, standard DCF and optimized DCF using the dynamic backoff method, whose target is to reduce collision and find the optimal transmission probability. The collision probability is calculated using (19) and the collision cost is defined as the average collision duration ($\sum_{m=1}^{M} P_{coll,m} T_{coll,m}$) normalized by the slot length. As shown in the figure, the collision probability of R-DCF is much smaller than original IEEE 802.11 DCF. With dynamic backoff, optimized DCF has a smaller collision probability than R-DCF in basic-access mode, whereas its collision probability is still higher than that of R-DCF in RTS/CTS mode. However, when the collision cost is measured, R-DCF enjoys a smaller collision cost even compared with the optimized DCF. Hence, a higher throughput in R-DCF is obtained in the sense of collision reduction.

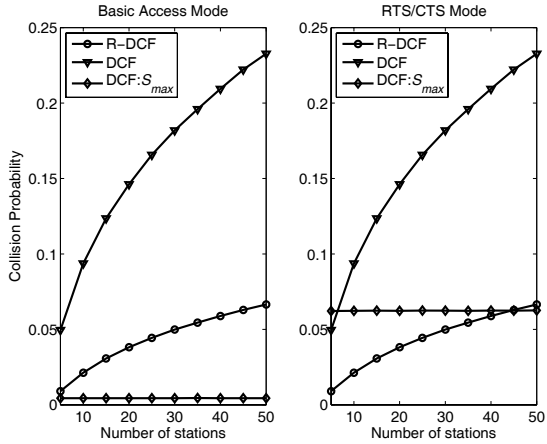

Fig. 6. Collision probability comparison.

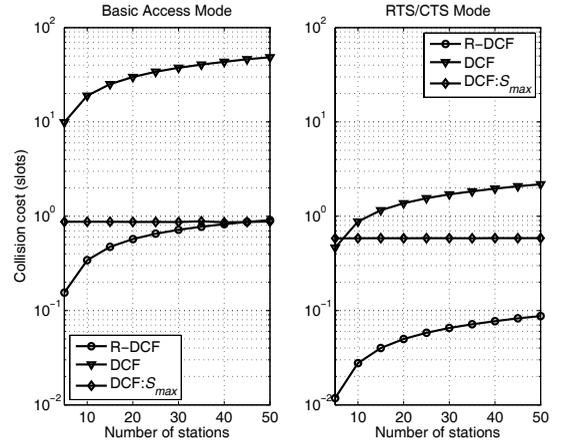

Fig. 7. Collision cost comparison.

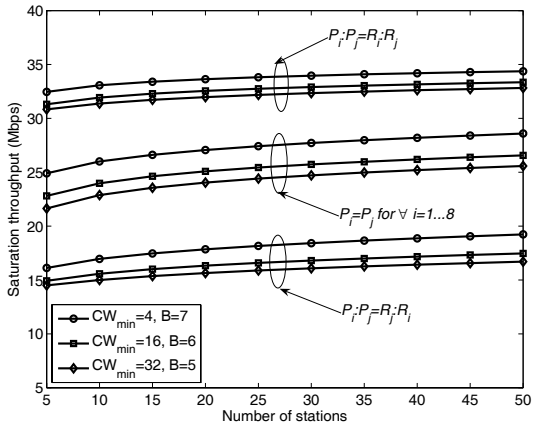

Fig. 8. Performance evaluation in basic-access mode.

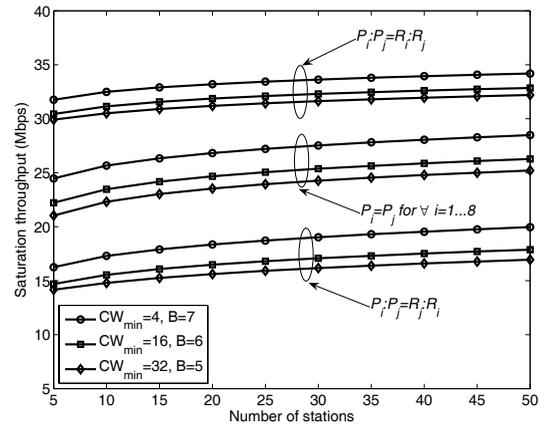

Fig. 9. Performance evaluation in RTS/CTS mode.

To further investigate the performance of R-DCF, in Fig. 8 and Fig. 9, we plot the throughput of R-DCF in the three different scenarios, namely, the equal, proportional and inversely proportional rate distributions, corresponding to $P_m = 1/8$, $P_i : P_j = R_i : R_j$ and $P_i : P_j = R_j : R_i$, respectively. In the proportional/inversely proportional rate distribution case, higher/lower rates are more likely to be adopted. In each case, we vary the minimum contention window size $CW_{min}$ and the maximum backoff stage $B$ pair ($CW_{min}$, $B$). As shown in the figures, unlike standard DCF where throughput drops with the increase of network size, R-

DCF enjoys a throughput increase with more stations present. Besides, with more transmissions at higher rates (proportional rate distribution), the system throughput is much higher. Interestingly, in each case a smaller $CW_{min}$ leads to a higher throughput in both basic access and RTS/CTS modes: the throughput with $CW_{min} = 4$ is higher than that of $CW_{min} = 16$, which is higher than that of $CW_{min} = 32$. This is because R-DCF has already reduced much collision through the rate-aware prioritization, and when the number of stations is not too large, the original contention window size results in too much idle time and consequently a lower throughput.

### 5.2 R-DCF in Fixed-rate WLANs

In the fixed-rate WLAN, a group-$m$ station uses rate $R_m$ for all its packet transmissions. For simplicity and without loss of generality, in our analysis and simulation we assume the number of stations in each group is the same, which means $N_m$ is the same for different $m$.

Fig. 10 and Fig. 11 demonstrate, respectively, the performance of R-DCF in the slow adaptation WLANs. For comparison, the performance of OAR/TXOP scheme, the performance achieved by differentiating MAC parameters and the maximal throughput of standard DCF are also plotted. Similar to the observations in the fast adaptation WLANs: 1) the throughput limit of standard DCF is quite low due to the presence of low-rate transmissions. 2)

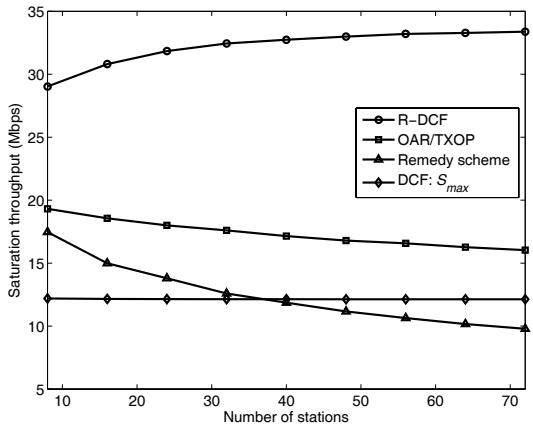

Fig. 10. Performance comparison in basic-access mode.

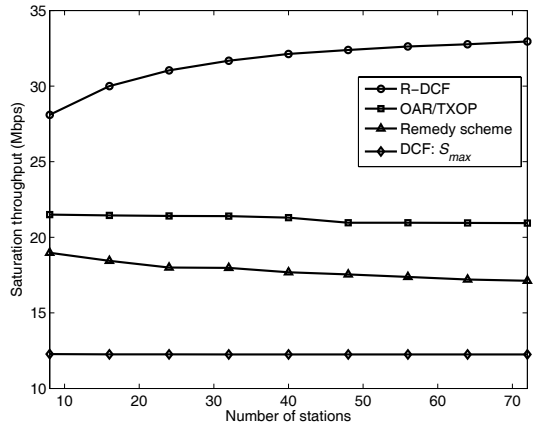

Fig. 11. Performance comparison in RTS/CTS mode.

The airtime fairness based OAR/TXOP schemes also improve the system throughput in the slow adaptation WLANs. Besides, the backoff based remedy scheme becomes effective in the slow adaptation scenario. Because stations' data rates do not change, using smaller contention window size for higher rate stations can indeed guarantee more channel access opportunity. In addition, R-DCF significantly outperforms other schemes in both basic access and RTS/CTS mechanism. Furthermore, the aggregate throughput of R-DCF in a slow adaptation WLAN is much higher than that of R-DCF in the fast adaptation WLAN, which can be seen by comparing Fig. 4 with Fig. 6 in basic-access mode, and Fig. 5 with Fig. 7 in RTS/CTS mode, respectively. The reason for this phenomenon is that: in a slow adaptation WLAN, the high rate stations always have an advantage over low rate stations and hence, have more chance to transmit because of our rate-aware prioritization. In contrast, in the fast adaptation WLAN, each station is possible to use low transmission rates and lose contention. Consequently, on average the channel access is suppressed. Therefore, the performance in the slow adaptation WLAN is better than that in the fast adaptation WLANs.

# 6 PERFORMANCE ENHANCEMENT OF R-DCF

In the previous sections, we present the framework of R-DCF and evaluate its performance with IEEE 802.11a MAC parameters (e.g. $CW_{min}$, $r$ etc.). However, these parameters are not optimal for R-DCF. In this section, we further investigate the optimization of system parameters in the fast adaptation WLAN with homogeneous users, aiming at maximizing the system throughput.

## 6.1 Maximizing Throughput of R-DCF

For a given network configuration (i.e. number of stations $N$, packet length $L$, available rates $R_m$'s and corresponding probability $P_m$'s), with homogeneous users the system throughput $S$ achieved by R-DCF is only a function of the zero backoff counter probability $\tau$ given by (10) and (17) – (19). Therefore, the maximal system throughput $S_{max}$ achievable by R-DCF can be derived by maximizing (10) with regard to $\tau$. In addition, since the stations are identi-cal in their rate distribution, each station also achieves its own maximum throughput through the maximization of (10). Let $\tau^*$ denote the optimal $\tau$ that achieves $S_{max}$. It can be calculated numerically by setting the first order derivative of $S$ to be zero. Furthermore, (16) shows that $\tau$ depends on $N$ and the backoff parameters (i.e. $CW_{min}$, $r$ and $B$). Note that given $\tau^*$, the corresponding zero backoff counter probability $f$ is already known from (17). Since $N$ is generally not controllable, a practical way to achieve $S_{max}$ in real systems is to adjust backoff parameters to achieve $\tau^*$. The above method is described mathematically in the following:

Step 1: Find the optimal $\tau^*$

$$\tau^* = \arg\max_{0 \le \tau \le 1}(S(\tau)) . \tag{24}$$

Step 2: Obtain optimal $CW_{min}$, $B$, and $r$ from the following equation.

$$\tau^* = \frac{2(1-rf)}{(1-rf)(CW_{min}+1)+CW_{min}f(r-1)\left[1-(rf)^B\right]} \tag{25}$$

subject to $CW_{min}, B \in \mathbb{N}$ and $r \ge 1$.

Note that by definition, the maximum backoff stage $B$ and minimum contention window $CW_{min}$ should be integers. Besides, to guarantee the stability, we require the backoff exponent $r \ge 1$.

In Fig. 12 we plot the system throughput as a function of the zero backoff counter probability $\tau$, given $N = 50$, $P_m = 1/8$ for all $m$ and $L = 2312$ bytes. As shown in the figure, the maximal throughput that can be achieved by R-DCF (denoted by circles) is much higher than the throughput achieved by using standard IEEE 802.11a parameters (denoted by diamonds). In addition, the optimal zero backoff counter probability $\tau^*$ is much larger than the probability obtained by using the standard IEEE 802.11a parameters. This is due to the fact that R-DCF has avoided much collision through its rate-aware prioritization. Hence the contention window used in standard IEEE 802.11a is too large to achieve the maximal throughput. To further evaluate the maximal throughput of R-DCF, in Fig. 13 we plot the maximal throughput $S(\tau^*)$ with different packet size. As shown in the figure, the maximal throughput of both access modes is almost independent of the network size. For $N = 10$ and $N = 50$, the throughput curves are quite close to each other. For basic access, the maximal throughput is almost independent of the packet size when it is larger than 1000 bytes, whereas for RTS/CTS access method, the maximal throughput increases with the packet size. Besides, when the packet size is larger than 1000 bytes, RTS/CTS access outperforms basic access even if there are only 10 stations.

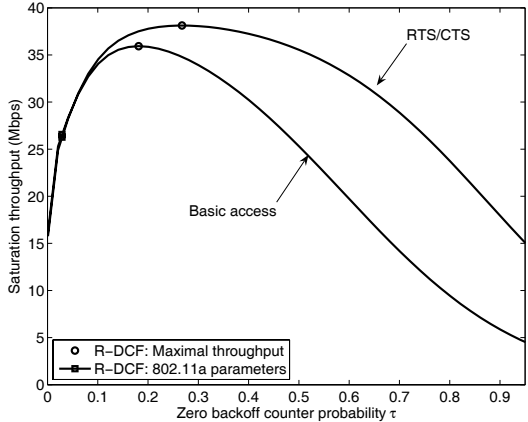

Fig. 12. Throughput vs. zero backoff probability: $N = 50$, $L = 2312$ bytes and $P_m = 1/8$ for all $m$.

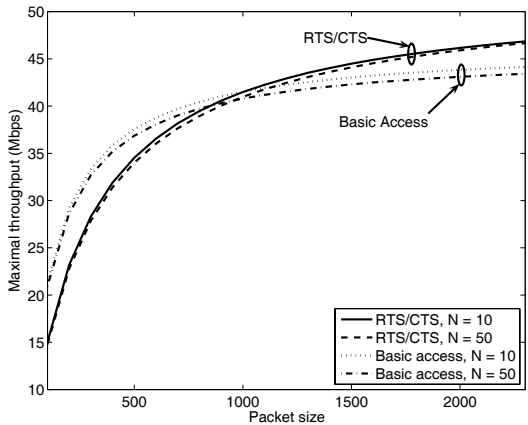

Fig.13. Maximal throughput of R-DCF with different packet size.

## 6.2 Offline Adaptive Backoff Method

The two-step method described in (24) and (25) guarantees maximal throughput performance. However, it involves solving nonlinear equations to find out the optimal zero backoff counter probability $\tau^*$ and the corresponding backoff parameters. In a wireless network, a sub-optimal but easy-to-implement method is often desirable. For the purpose of easy implementation in real systems, in this subsection we propose a simplified adaptive backoff method to approach the maximum throughput at low complexity. Note that (24) can be easily solved by numerical methods. Thus, we focus on how to find the corresponding backoff parameters in (25), which requires solving one nonlinear equation with three unknowns. The principles of our offline adaptive backoff strategy are as follows:

1 As pointed out in [20], the maximum backoff stage $B$

## TABLE 3
OPTIMAL $\tau^*$ AND BACKOFF PARAMETERS

| N | Basic Access Method | | | | RTS/CTS Access Method | | | |
|---|---|---|---|---|---|---|---|---|
| | $\tau^*$ | $r_{opt}$ | $r_{app}$ | CW | $\tau^*$ | $r_{opt}$ | $r_{app}$ | CW |
| 5 | 1 | 1 | 1 | 1 | 1 | 1 | 1 | 1 |
| 10 | 0.8592 | 1 | 1 | 1 | 1 | 1 | 1 | 1 |
| 15 | 0.5861 | 1 | 1 | 1 | 0.8505 | 1 | 1 | 1 |
| 20 | 0.4445 | 1.117 | 1.2 | 2 | 0.6488 | 1 | 1 | 1 |
| 30 | 0.2995 | 1.223 | 1.2 | 2 | 0.4399 | 1.109 | 1.1 | 2 |
| 40 | 0.2258 | 1.298 | 1.3 | 2 | 0.3328 | 1.179 | 1.2 | 2 |
| 50 | 0.1812 | 1.356 | 1.3 | 2 | 0.2676 | 1.233 | 1.2 | 2 |
| 60 | 0.1513 | 1.404 | 1.4 | 2 | 0.2238 | 1.278 | 1.3 | 2 |
| 70 | 0.1299 | 1.445 | 1.4 | 2 | 0.1923 | 1.315 | 1.3 | 2 |
| 80 | 0.1138 | 1.481 | 1.5 | 2 | 0.1685 | 1.349 | 1.4 | 2 |
| 90 | 0.1012 | 1.514 | 1.5 | 2 | 0.1500 | 1.378 | 1.4 | 2 |
| 100 | 0.0912 | 1.543 | 1.5 | 2 | 0.1352 | 1.405 | 1.4 | 2 |

only has marginal impact on system performance when the network size is not too large. Thus, fix $B$ to be 6 as what is used in 802.11a standard.

2 To avoid overhead of accurate estimation of network size, keep $CW_{min}$ independent of the network size. Thus, given different $\tau^*$ corresponding to different $N$, try to fix $CW_{min}$ when solving (25).

3 Given $B = 6$ and a designated $CW_{min}$, solve (25) to find the optimal backoff exponent $r$.

4 These parameters can be pre-calculated and stored as an offline table in a station. At runtime, a station only needs to estimate roughly the network size (estimation algorithm can be found in [9], [10], [23]) and choose suitable $(CW_{min}, r)$ accordingly.

As an example, given $P_m =1/8$ and $L =1028$ bytes, we solve the optimal backoff parameters using the above procedure. The optimal $\tau^*$ and corresponding system parameters are listed in Table III. From the table we can see that when the network size is not large ( $N < 20$ for basic access and $N < 25$ for RTS/CTS access), $\tau^*$ is or is close to 1, which implies that it is preferable to transmit immediately based on our prioritized collision avoidance without random backoff. Thus, we set $CW_{min}=1$, and $r =1$. When $N$ increases, we always fix $CW_{min}=2$ and solve (25) to get optimal $r$. From our calculation, the optimal $r$ is typically within $1.2 \sim 1.5$ for different $N$. To further reduce the runtime complexity, we use the approximated backoff exponent $r_{app}$ instead of $r_{opt}$ in the simulations. As shown in Table III, $r_{app}$ does not vary significantly with $N$, which further reduces the requirement of accurate measurement of network size $N$. Note that in R-DCF $CW_{min}$ and $r$ are smaller than in DCF. This is because R-DCF has already avoided much collision by its rate-aware prioritization and smaller parameters make stations more aggressive for channel contention. Consequently, high rate links

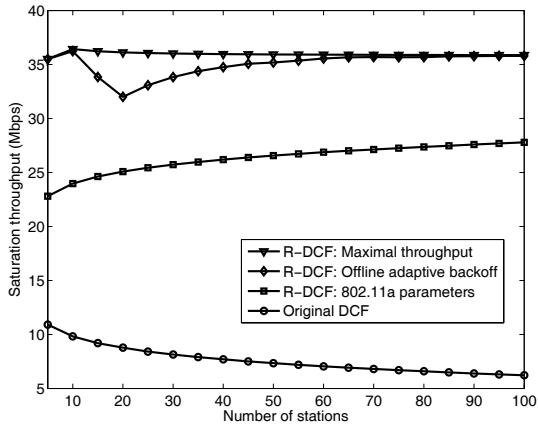

Fig. 14. R-DCF with offline adaptive backoff: basic access mode

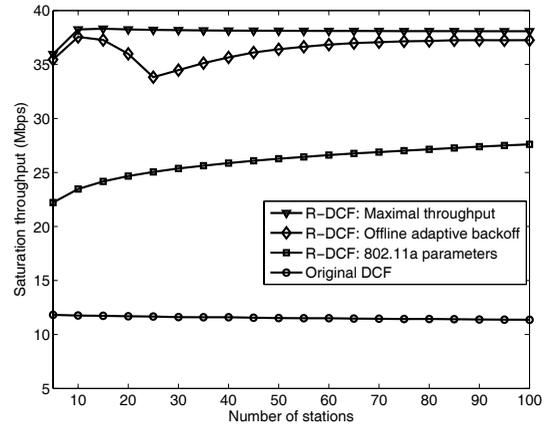

Fig. 15. R-DCF with offline adaptive backoff: RTS/CTS mode.

are more likely to win.

Fig. 10 and 11 compare (i) the maximal throughput R-DCF, (ii) the throughput of R-DCF obtained by the offline adaptive backoff method, (iii) R-DCF with standard IEEE 802.11a backoff parameters, and (iv) original 802.11 DCF. As shown in the figures, the performance of the offline adaptive backoff method approaches the maximal throughput quickly with the increase of network size. Besides, we observe a discrepancy between the theoretical upper bound of R-DCF and Enhanced R-DCF when the $N$ is from 10 to 30. This is due to the approximation of the integer contention window size and backoff counter value, which we use to achieve the optimal $\tau^*$. When these values are small, the relative error from approximation makes the algorithm less effective. As shown in the figures, the simulations converge to the theoretical results quickly as the network size increases.

In summary, given the channel fading characteristics, the maximal throughput and corresponding parameters can be calculated using the two-step maximization method. If runtime complexity is not a major concern, online maximization as described in (22) and (23) guarantees the optimal throughput performance. On the other hand, based on a rough estimation of network size and using the approximate system parameters, the offline adaptive backoff method is able to provide a close-to-optimal performance at low runtime computational cost.

## 7 CONCLUSION

In this paper, we have proposed and analyzed a novel R-DCF protocol to effectively exploit multi-user diversity in a fully distributed manner in multi-rate IEEE 802.11 WLANs. The proposed R-DCF protocol drastically improves system performance in the following senses: (i) system throughput is greatly enhanced, for mobile stations are more likely to access the channel at their peak rates. (ii) Channel is more efficiently utilized, because low-rate transmissions no longer jeopardize the channel access of high-rate links. (iii) Collision cost is dramatically reduced, thanks to the rate-aware prioritization strategy. The proposed R-DCF protocol shows notably superiority compared with existing schemes in both basic access and RTS/CTS mechanism. Based on the analysis of R-DCF, we further maximize the throughput of R-DCF by optimizing the exponential backoff. For easy implementation, an offline adaptive backoff method is proposed to approach the maximal throughput with low runtime cost. Numerical results show that using the offline adaptive backoff method, R-DCF achieves further throughput improvement of 29% in the basic access mode and 35% in the RTS/CTS access mode.

The adaptive backoff method is designed under the homogeneous users' assumption. In this case each user achieves the same maximal throughput when the system is optimized and fairness among users is automatically guaranteed. In our future work, we will extend the similar methodology to WLANs with heterogeneous users, where fairness among users becomes an optimization constraint.

## ACKNOWLEDGMENT

This work is supported in part by the Competitive Earmarked Research Grant (Project Number 418506) established under the University Grant Committee of Hong Kong, and the Direct Grant (Project Number 2050370) of the Chinese University of Hong Kong.